\documentclass[pre,showpacs,preprint]{revtex4}
\usepackage{amssymb,graphicx,amsbsy,amsmath}

\begin{document}
\title{Upper transition point for percolation on the enhanced binary tree:
A sharpened lower bound}
\author{Seung Ki Baek}
\affiliation{ School of Physics, Korea Institute for Advanced Study, Seoul
130-722, Korea}
\email[Corresponding author, E-mail: ]{seungki@kias.re.kr}

\begin{abstract}
Hyperbolic structures are obtained by tiling a hyperbolic surface with
negative Gaussian curvature. These structures
generally exhibit two percolation transitions:
a system-wide connection can be established
at a certain occupation probability $p=p_{c1}$
and there emerges a unique giant cluster at $p_{c2} > p_{c1}$.
There have been debates about locating the upper transition point of a
prototypical hyperbolic structure called the enhanced binary tree (EBT),
which is constructed by adding loops to a binary tree.
This work presents its lower bound as $p_{c2} \gtrsim 0.55$ by using
phenomenological renormalization-group methods and discusses some solvable
models related to the EBT.
\end{abstract}

\maketitle

\section{Introduction}

\begin{figure}
\includegraphics[width=0.20\textwidth]{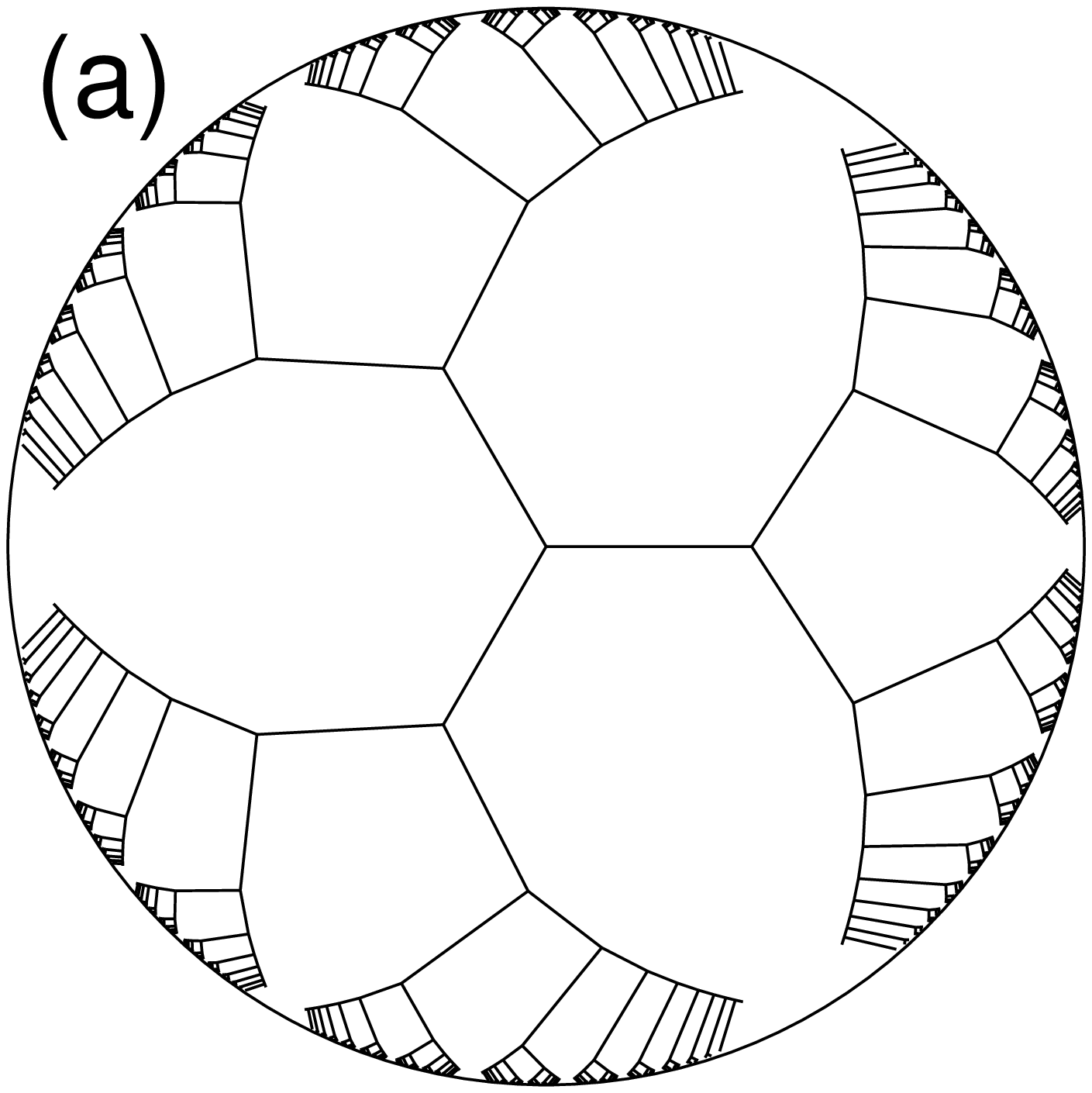}
\includegraphics[width=0.20\textwidth]{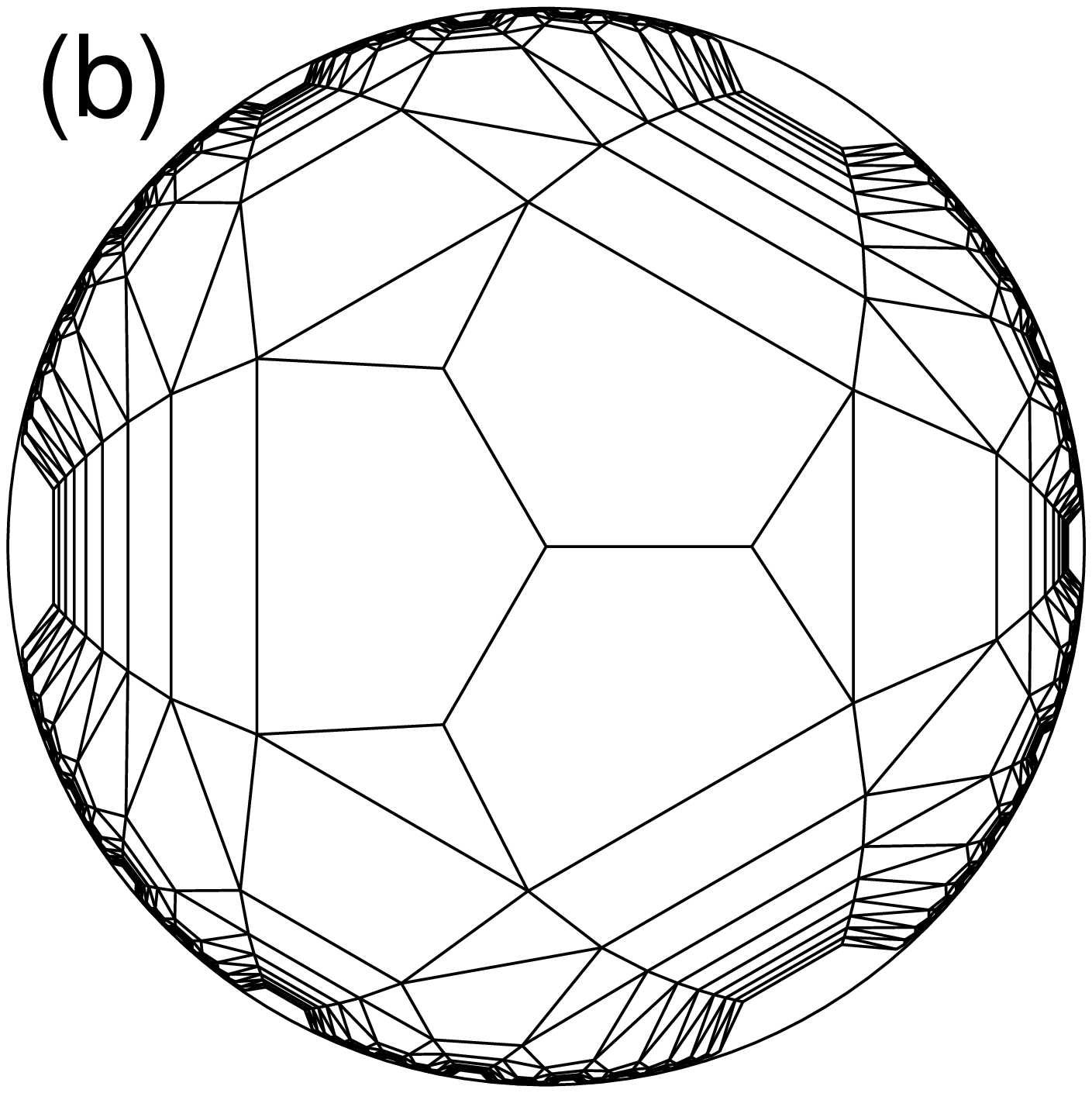}
\includegraphics[width=0.33\textwidth]{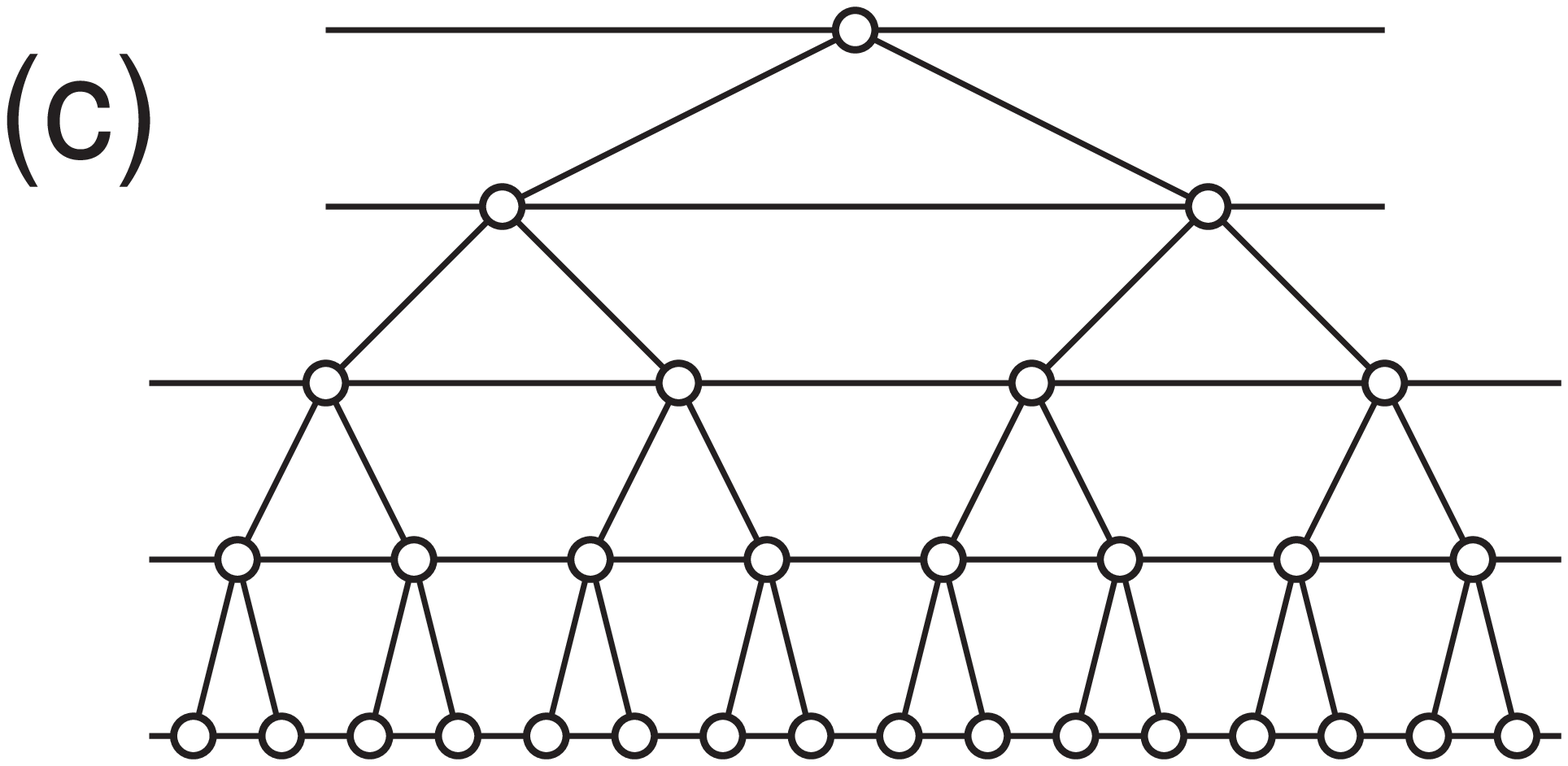}
\caption{(a) Schematic representations of a simple binary tree and (b) of
the EBT derived from (a). These are drawn on the
Poincar{\'e} disk, where the circular boundary indicates points at infinity.
(c) A part of the EBT.}
\label{fig:ebt}
\end{figure}

Percolation has been one of the most popular model systems in
statistical physics and it still remains as an active research area. For a
classical introduction, one may refer to Refs.~\cite{stauffer,grimmett}.
One recent observation in this field is that there generally occur two
percolation transitions if the size of a given system expands exponentially
fast as its length scale grows. For example, the size of a binary tree
increases as $N \sim 2^L$ with the number of layers $L$
[Fig.~\ref{fig:ebt}(a)].
At the first percolation point $p^{\rm tree}_{c1} = 1/2$, it becomes
possible to establish a global connection, and the resulting cluster size
scales linearly with $L$.
However, this cluster size is still negligible compared with $N$
since $L / 2^L \rightarrow 0$ as $L$ increases. We find the
largest cluster size $s_1$ comparable to $N$ only at $p=1$, which
determines $p^{\rm tree}_{c2}=1$. This tree in fact belongs to a
category called hyperbolic lattices, obtained by tessellating a hyperbolic
surface appearing in hyperbolic geometry.
Since such double percolation transitions in hyperbolic structures were
revealed by numerical calculations~\cite{perc}, there have been debates
about locating the upper transition point~\cite{ebt,comment,an,bnd},
particularly by dealing with a prototypical lattice model called the enhanced
binary tree (EBT). This structure is not a tree in itself, but is derived from
the binary tree by connecting vertices on the same layer horizontally
[Fig.~\ref{fig:ebt}(b)]. It thus describes spreading along a branching
structure with possible horizontal transfer.
While the duality relation implies $p_{c2} =
0.564(1)$~\cite{ebt},
we have obtained $p_{c2} \approx 0.5$ by utilizing
a simple extrapolation of the largest cluster size
$s_1 \sim N^{-\phi}$, which is correct for a tree~\cite{perc}.
This looks also consistent with the
observation of $s_2/s_1$ where $s_2$ is the size of the second largest
cluster~\cite{comment}.
We have even tried to
explain this estimate $p_{c2} = 1/2$ analytically in combination with
numerical observations and approximate renormalization-group
methods~\cite{an,bnd}, pointing out that the duality argument does not have
a solid mathematical ground here.

Recently, Ref.~\cite{ziff} revisited this issue by calculating
the crossing probability. According to the conformal field
theory~\cite{cardy}, the crossing probability for a unit disk whose boundary
is divided at four points $z_1, z_2, z_3$ and $z_4$ is given by
\begin{equation}
R = \frac{\Gamma(\frac{2}{3})}{\Gamma(\frac{4}{3})\Gamma(\frac{1}{3})}
\eta^{\frac{1}{3}} ~_2F_1(\frac{1}{3}, \frac{2}{3}; \frac{4}{3}, \eta),
\label{eq:cardy}
\end{equation}
with the gamma function $\Gamma$, the hypergeometric function $_2F_1$, and
the cross ratio
\[
\eta = \frac{(z_1-z_2)(z_3-z_4)}{(z_1-z_3)(z_2-z_4)}.
\]
If the boundary is divided into four equal pieces, e.g., $z_1=-1$, $z_2=-i$,
$z_3=1$, and $z_4=i$, the cross ratio becomes $\eta = 1/2$ and we
immediately find $R=1/2$ (see, e.g., Refs.~\cite{kleban,rietman} on the
connection between the hyperbolic geometry and the conformal field theory).
Such a point where $R=1/2$ is denoted as a duality point in
Ref.~\cite{ziff}.
For each of several hyperbolic structures considered there,
they have numerically calculated $R(p)$ by dividing the
boundary into four equal intervals.
Then by extrapolating $R(p)$ to the large-size limit at the inflection
point, Ref.~\cite{ziff} suggests that the limiting tangent line
gives an upper bound of $p_{c1}$ and a lower bound of $p_{c2}$.
A notable point is that the slope of the line converges to a finite value,
which clearly differs from the two-dimensional (2D) results.
This method yields $p_{c2} \ge 0.503$ for the EBT, questioning the validity
of the claim that $p_{c2}=1/2$.
They have also estimated $p_{c2}$ as 
$0.564(10)$ by extrapolating the value of $p$ where 
$R(p) = 1 - \epsilon$ with $\epsilon \ll 1$ as growing the system size,
which is consistent with the estimate in Ref.~\cite{ebt}.
In this paper, equipped with better analytic tools than before, we too reach a
conclusion that $p_{c2}$ is indeed larger than $1/2$. Our new
lower bound, $p_{c2} \gtrsim 0.55$, is obtained by transfer-matrix
calculations for percolation and includes the lower bound in Ref.~\cite{ziff}.

This paper is organized as follows. In Sec.~\ref{sec:solv}, we consider two
solvable models. Even though both of them have the trivial transition point
$p_{c2}=1$, this consideration gives an insight about
percolation in the EBT. Then in Sec.~\ref{sec:ebt}, we deal with the EBT in two
different ways: one is the block-cell transformation and the other is the
transfer-matrix method. Both of them lead to $p_{c2} > 1/2$ but the latter
gives a sharper bound. We then conclude this work by reexamining our
previous estimate in Sec.~\ref{sec:con}.

\section{Solvable models}
\label{sec:solv}
\subsection{Ternary tree}
\begin{figure}
\includegraphics[width=0.27\textwidth]{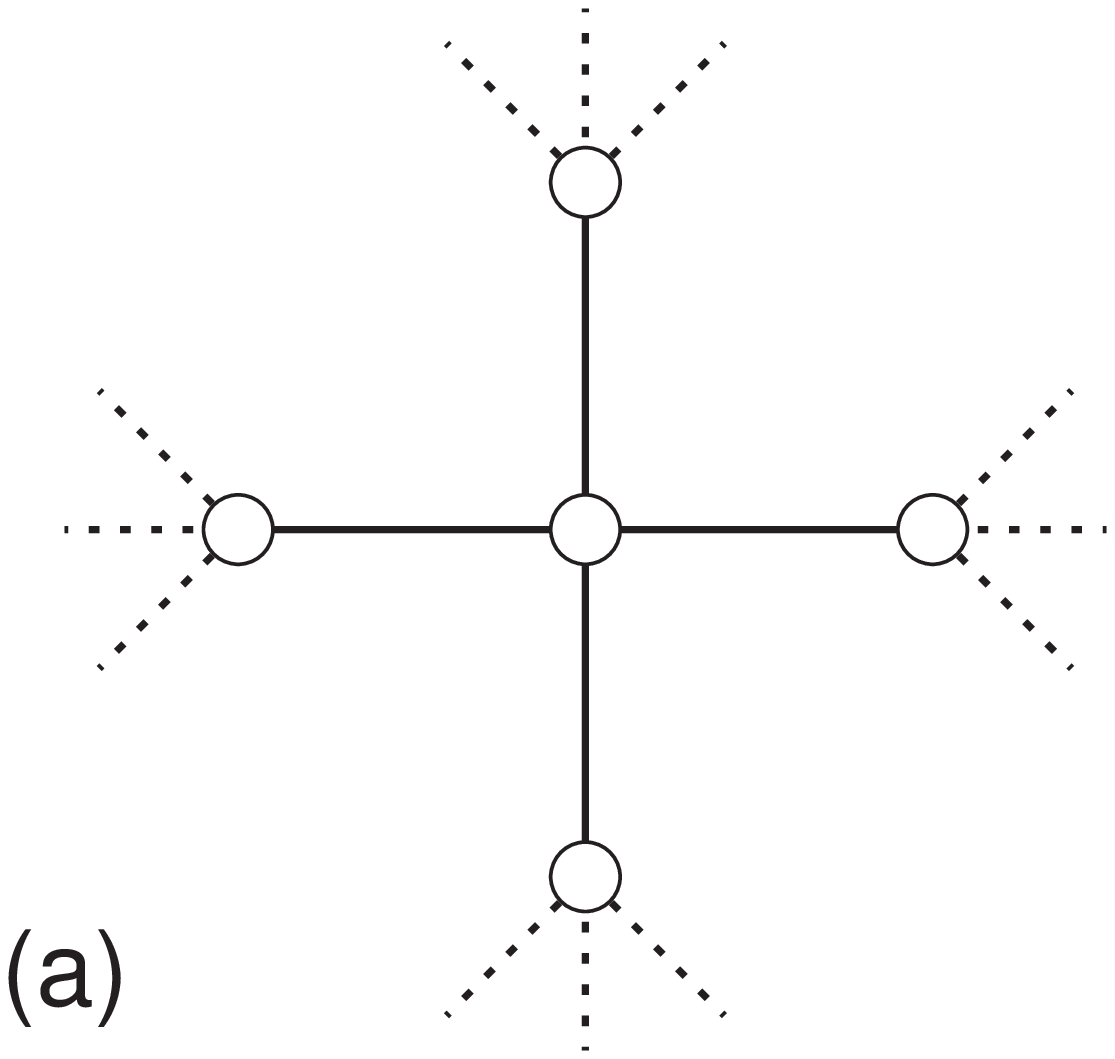}
\includegraphics[width=0.45\textwidth]{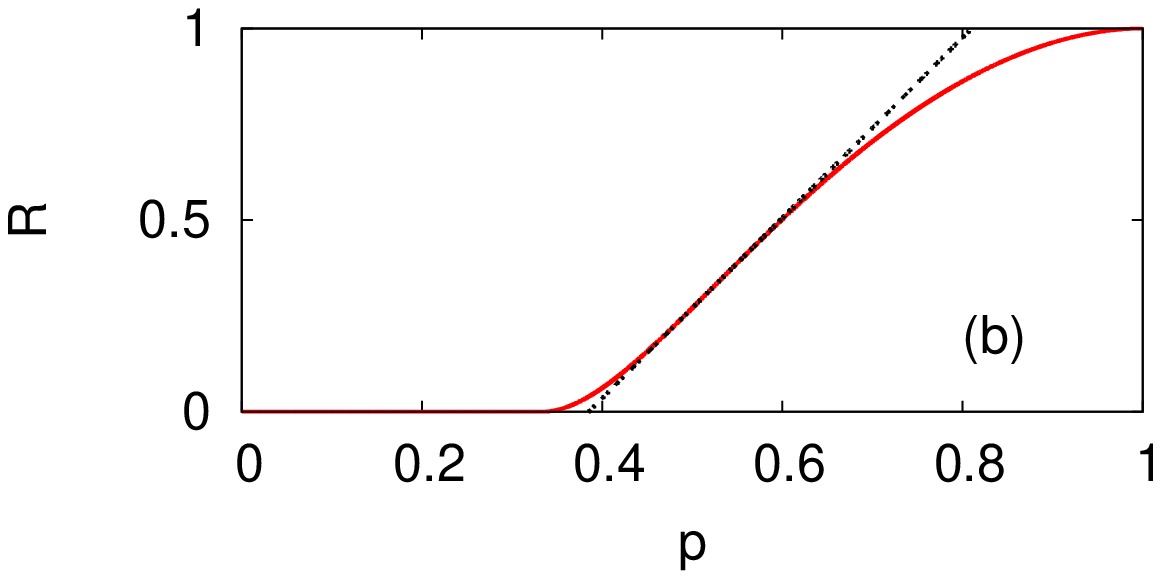}
\caption{(Color online)
(a) Schematic representation of a ternary tree. (b) Crossing
probability $R$ given by Eq.~(\ref{eq:tre}) with a tangent line at the
inflection point $p \approx 0.524~032$ (dotted line).}
\label{fig:tre}
\end{figure}
A tree with coordination number $g=4$ is the simplest example to
calculate the crossing probability. The coordination number is chosen to
provide the structure with natural four-fold symmetry. The center node has
four branches [solid lines in Fig.~\ref{fig:tre}(a)], each of which leads to
a tree with branching ratio $b=g-1=3$ [dotted lines in Fig.~\ref{fig:tre}(a)].
Inside each tree, the probability $\psi$ to connect the top
to the boundary can be described by the Galton-Watson process~\cite{will},
where the extinction probability $w$ can be identified with $1-\psi$.
The number of offsprings $k$ for each node is chosen from a binomial
distribution $B(3,p)$, where $p$ is the occupation probability of each bond.
The generating function is readily obtained as $\phi(s) \equiv
\sum_{k=0}^{\infty} s^k \binom{3}{k} p^k (1-p)^{3-k} = (1-p+sp)^3$.
The extinction probability is then given by a solution of the equation
$\phi(w) = w$~\cite{will}, which is
\[
w = 
\left\{
\begin{array}{lc}
1 & \mbox{if~}0\le p \le \frac{1}{3}\\
\frac{-3p^2 + 2p^3 + \sqrt{4p^3 - 3p^4}}{2p^3} & \mbox{if~}\frac{1}{3} \le p
\le 1,
\end{array}
\right.
\]
or, equivalently,
\[
\psi = 
\left\{
\begin{array}{lc}
0 & \mbox{if~}0 \le p \le \frac{1}{3}\\
\frac{3p^2 - \sqrt{4p^3 - 3p^4}}{2p^3} & \mbox{if~}\frac{1}{3} \le p \le 1.
\end{array}
\right.
\]
If we also consider the probability to connect two opposite branches of the
central cross in Fig.~\ref{fig:tre}(a), we get the crossing probability as
\begin{equation}
R(p) = 2 p^2 (1-p)^2 \psi^2 + 4 p^3 (1-p) \psi^2 + p^4 [
2\psi^2(1-\psi)^2 + 4\psi^3(1-\psi) + \psi^4],
\label{eq:tre}
\end{equation}
where $2p^2 (1-p)^2$, $4p^3 (1-p)$, and $p^4$ describe the connecting
configurations of the central cross and the other $\psi$-dependent parts
describe configurations of the trees attached to the cross.
One should note that we have considered crossing in either direction, while
it is in only one given direction in Eq.~(\ref{eq:cardy}) and
Ref.~\cite{ziff}. This difference in the definition of crossing will not
change any essential behavior, however.
We plot the result in Fig.~\ref{fig:tre}(b), and
an interesting point is that the slope of this function is finite everywhere
between $p_{c1}=1/3$ and $p_{c2}=1$,
in accordance with the numerical analysis in
Ref.~\cite{ziff}.
Note that this is markedly different from the 2D
percolation where the slope diverges at the critical point.
We also see that the tangent line at the inflection point $p \approx 0.524~032$
does give a lower bound of $p_{c2}$ as well as an upper bound of $p_{c1}$ as
suggested in Ref.~\cite{ziff}.

\subsection{Binary tree with a ring at the boundary}
\begin{figure}
\includegraphics[width=0.15\textwidth]{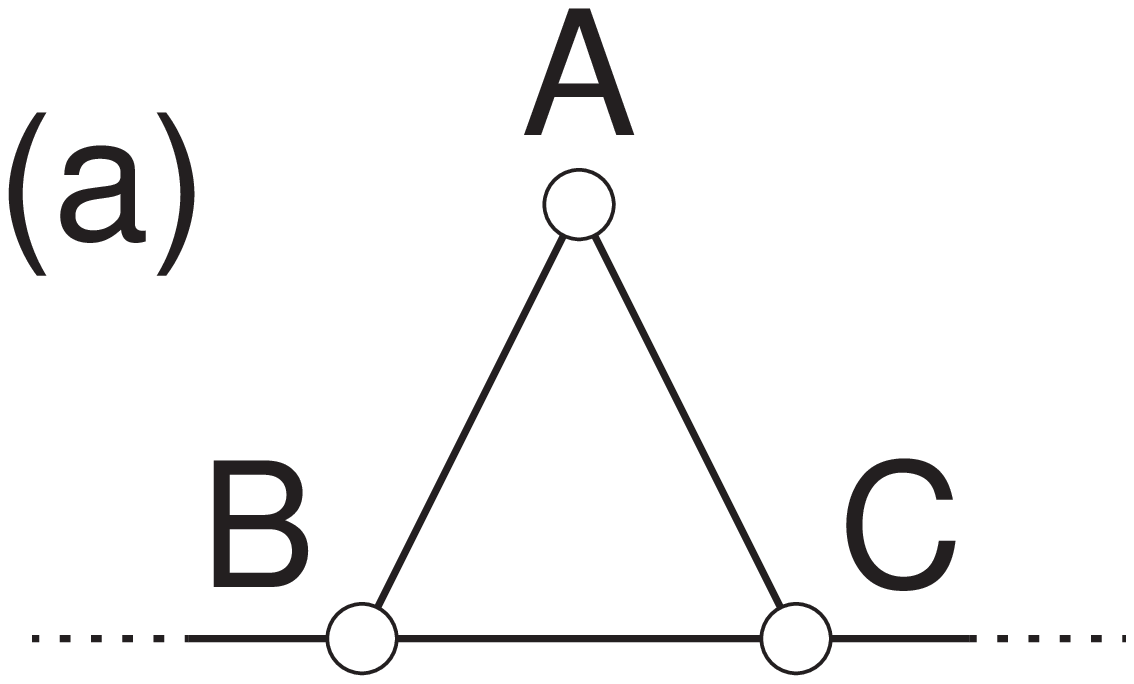}
\includegraphics[width=0.25\textwidth]{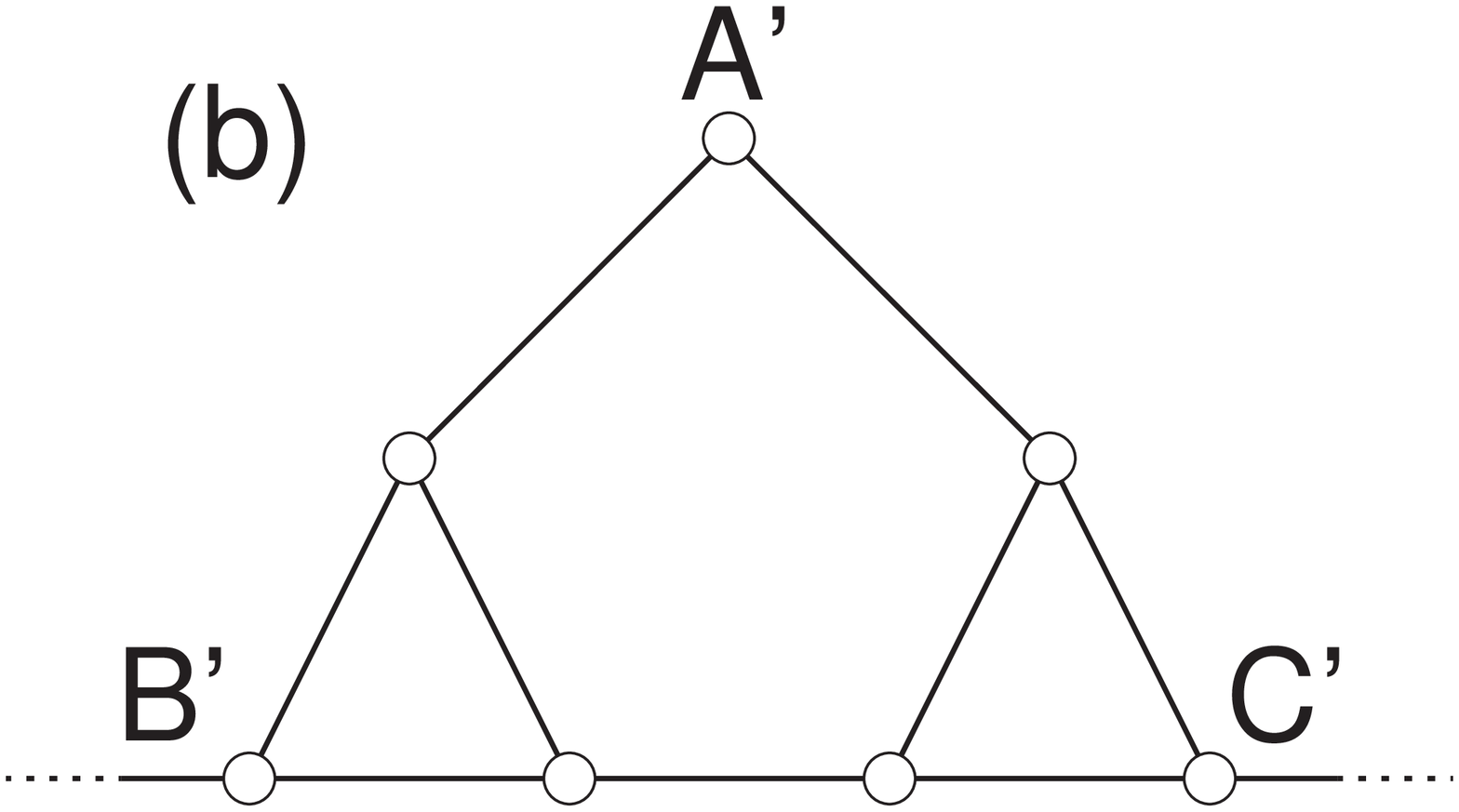}\\
\includegraphics[width=0.45\textwidth]{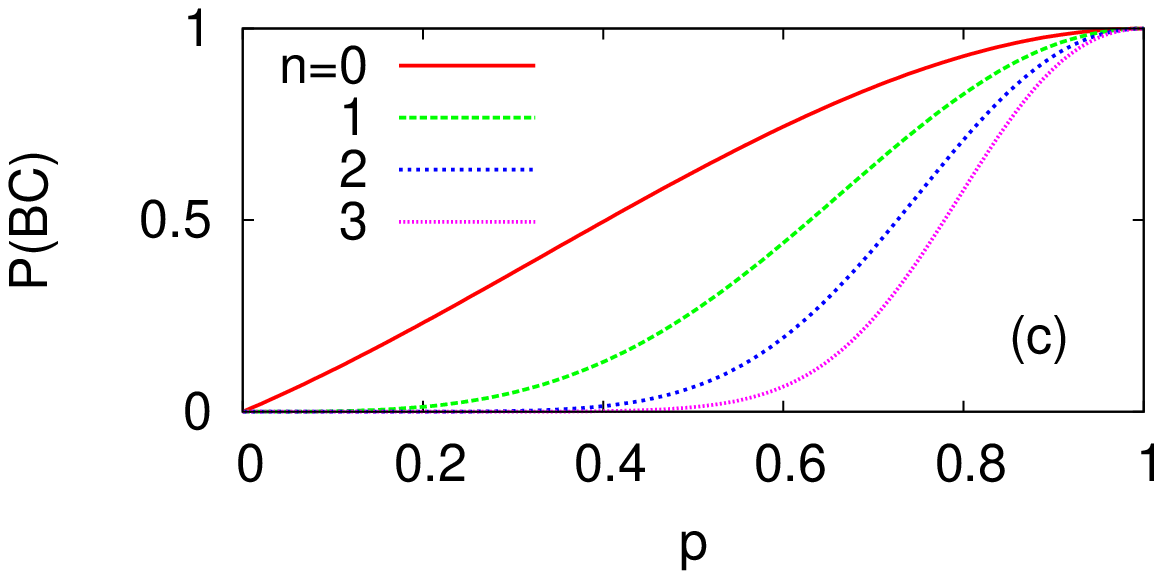}
\includegraphics[width=0.45\textwidth]{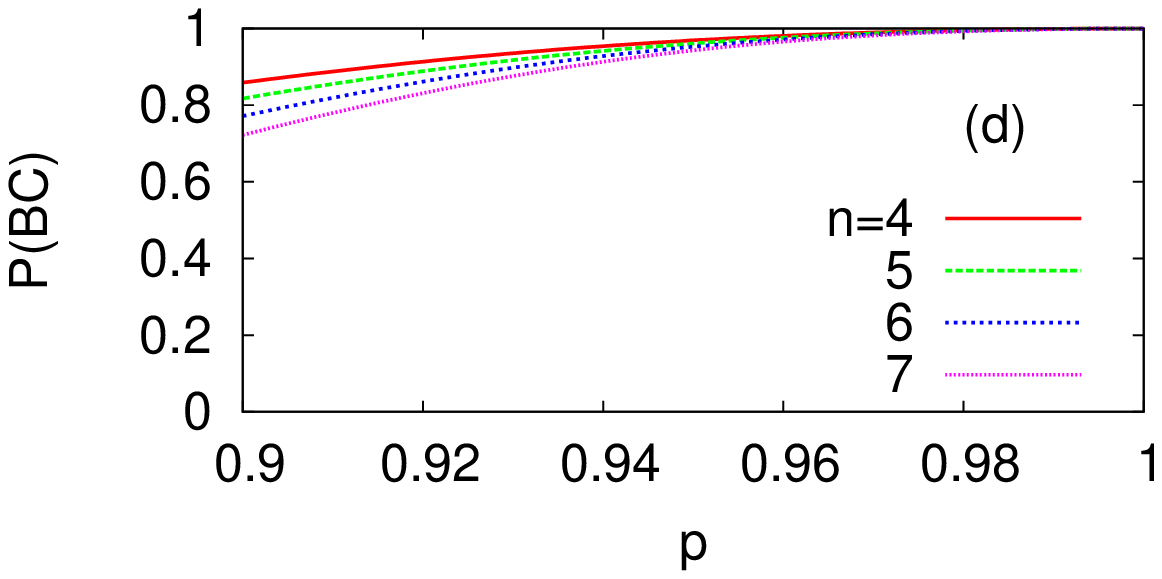}
\caption{(Color online)
Binary tree with a ring at the boundary. (a),(b) Probabilities of
connection in $\triangle ABC$ determine those in $\triangle A'B'C'$,
resulting in the recursion Eq.~(\ref{eq:recur}). (c) The probability of
connection between two farthest points on the boundary, $P(BC)$, when the
recursion equation is iterated $n$ times. (d) The same quantity for $p \ge
0.9$ with $n=4,\ldots,7$.}
\label{fig:ring}
\end{figure}
Let us add loops to a tree by attaching a ring along
the boundary points. This may be regarded as a first step toward making the
EBT, and even this single ring can introduce a large number of loops into the
system.
We first focus on the smallest triangle touching the
boundary and denote it as $\triangle ABC$ [Fig.~\ref{fig:ring}(a)].
Following Ref.~\cite{cell}, we define $P(ABC)$ as the
probability that all the three points are connected to one another, while
$P(\bar{A}\bar{B}\bar{C})$ as the probability that there is no
connection among them. In addition, $P(\bar{A}BC)$ means the probability
that $B$ and $C$ are connected but $A$ is not. One can also define
$P(A\bar{B}C)$ and $P(AB\bar{C})$ in the same way. These cover the
whole possibilities by
\[
P(ABC) + P(\bar{A}\bar{B}\bar{C}) + P(\bar{A}BC) + 
P(A\bar{B}C) + P(AB\bar{C}) = 1.
\]
Considering a larger triangle $\triangle A'B'C'$ containing the two smallest
triangles [Fig.~\ref{fig:ring}(b)],
we find that it is possible to express the five probabilities of 
$\triangle A'B'C'$ with those of $\triangle ABC$. Note that the left-right
symmetry is preserved by this transformation so that we have three independent
variables $x \equiv P(ABC)$, $y \equiv P(\bar{A}BC)$, and
$z \equiv P(A\bar{B}C)$. After some algebra, the transformation turns out to be
\begin{eqnarray}
x' &=& p^2 (z^2 + 2xz + 2xy - 2px^2 + 3x^2),\nonumber\\
y' &=& p (y - px + x)^2,\label{eq:recur}\\
z' &=& -p (pz^2 - pyz + p^2xz + pxz - z + pxy - p^2x^2 + 2px^2 -
x),\nonumber
\end{eqnarray}
where the prime is in order to indicate probabilities for $\triangle A'B'C'$.
The initial condition is given by counting the possibilities in $\triangle
ABC$ as
\begin{eqnarray*}
x &=& p^3 + 3p^2(1-p),\\
y &=& p(1-p)^2,\\
z &=& p(1-p)^2.
\end{eqnarray*}
The quantity of interest is $P(BC) = P(ABC) + P(\bar{A}BC) = x+y$, and this
can be obtained exactly at every iteration step [Fig.~\ref{fig:ring}(c)].
Note that this quantity is closely related to the crossing probability since
it measures the chance for a boundary point to connect to another boundary
point far away, which is possibly achieved through the inner part of the
system.
When this transformation is iterated, we observe that $P(BC)$ eventually
vanishes except at $p=1$ [Fig.~\ref{fig:ring}(c)], so we conclude that the
added loops are not enough to make $p_{c2}$ nontrivial. However, it is
notable that the convergence is so slow that it is hard to determine
$p_{c2}$ by naive extrapolation [Fig.~\ref{fig:ring}(d)].

\section{Enhanced binary tree}
\label{sec:ebt}
\subsection{Block-cell transformation}
\begin{figure}
\includegraphics[width=0.30\textwidth]{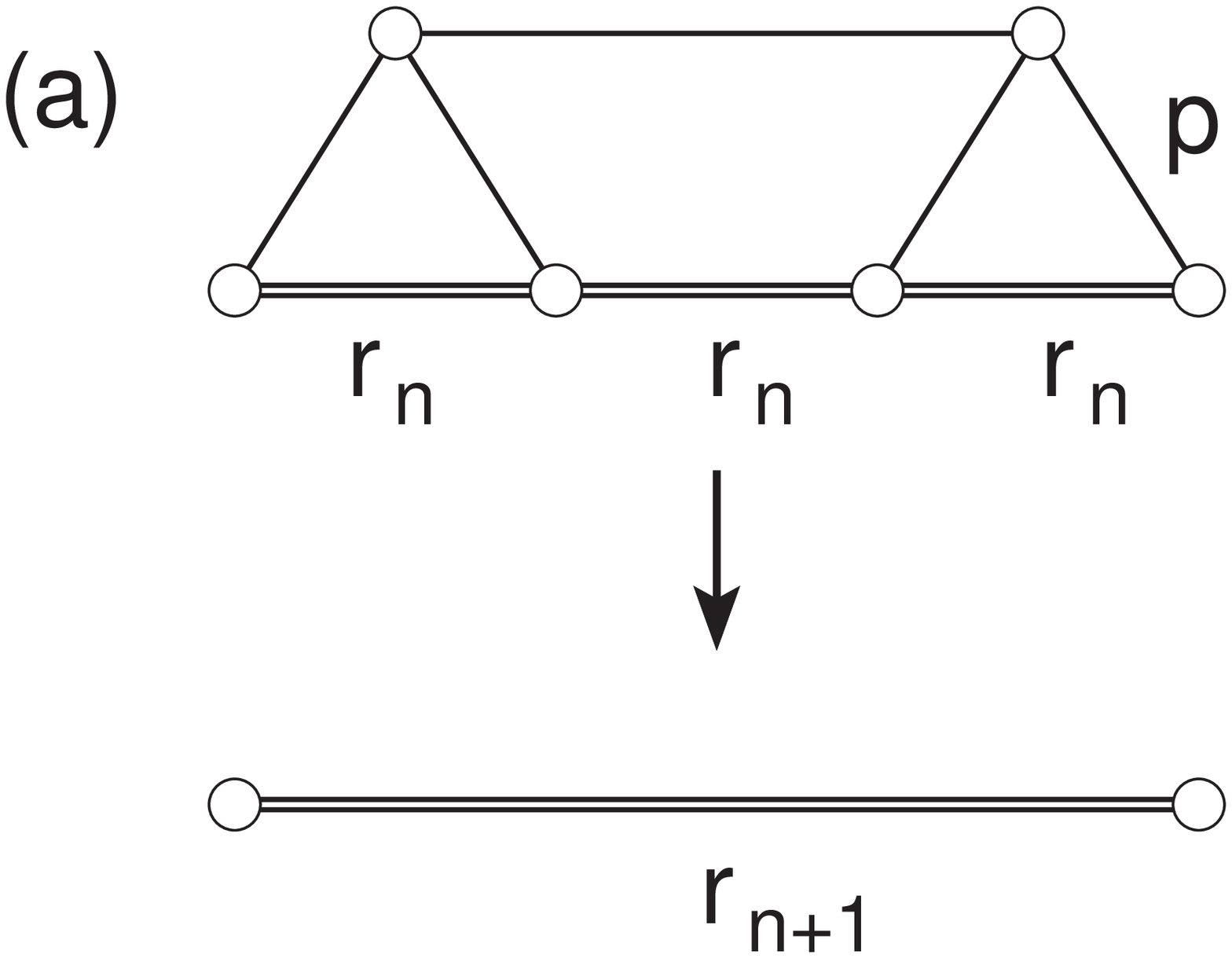}
\includegraphics[width=0.55\textwidth]{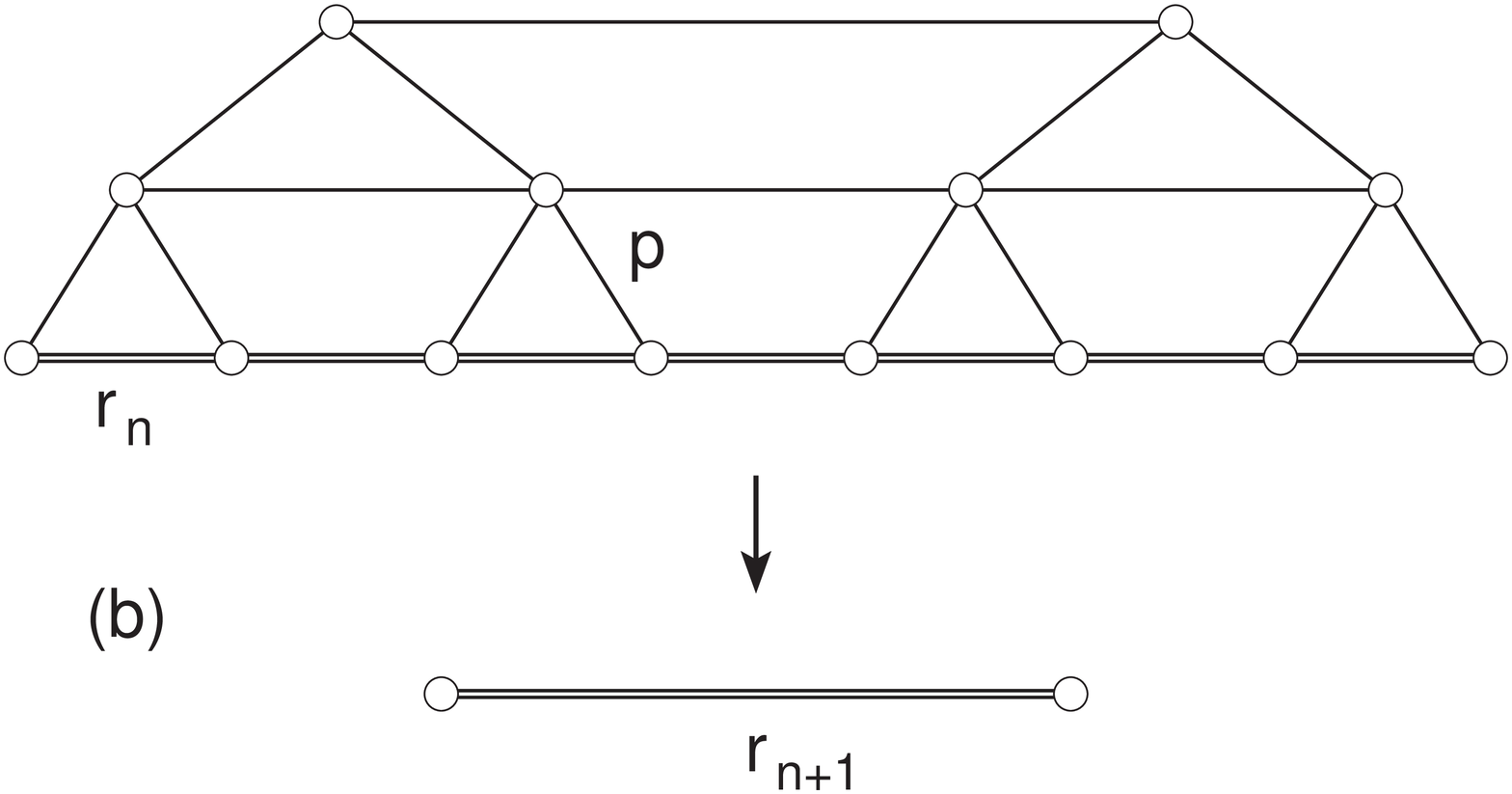}
\includegraphics[width=0.45\textwidth]{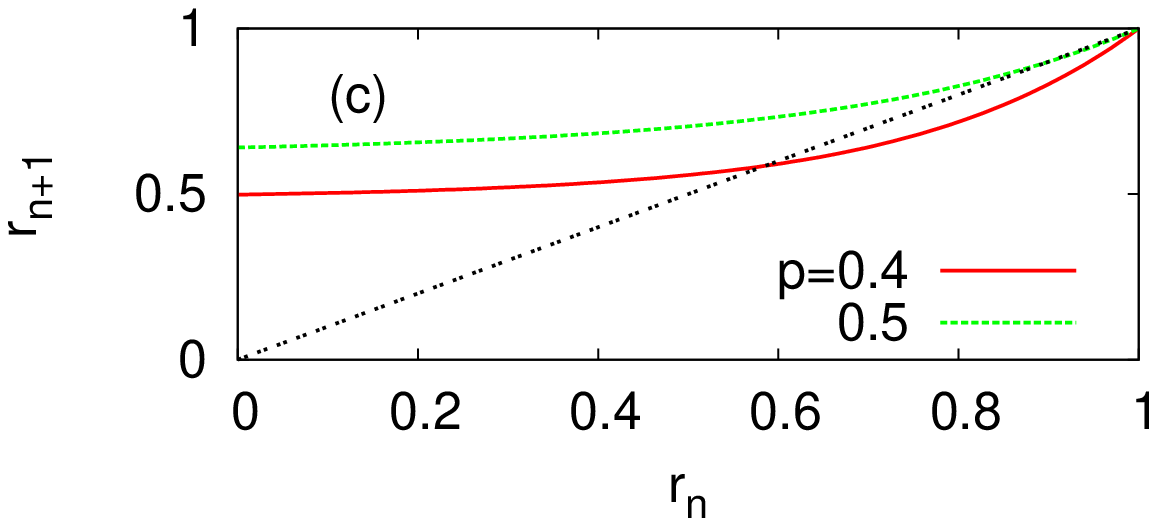}
\includegraphics[width=0.45\textwidth]{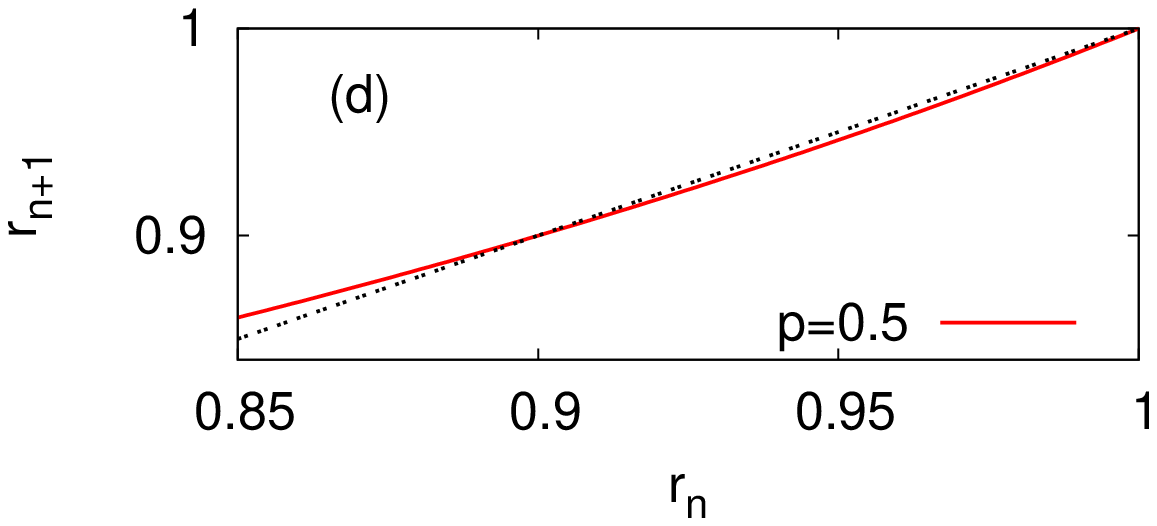}
\caption{(Color online)
(a) Block-cell transformation for eight bonds, where $r_n$ is the
renormalized connection probability at the $n$-th iteration step,
represented by double lines. (b)
Transformation for 23 bonds and (c) its application at $p=0.4$ and $0.5$.
The dotted line indicates a slope of $1$.
(d) A zoomed view at $p=0.5$ shows that this $p$ is still below $p_{c2}$. }
\label{fig:block}
\end{figure}
The block-cell transformation shown in Fig.~\ref{fig:block}(a) was
introduced to get a lower bound of $p_{c2}$ in Ref.~\cite{bnd}. It yields a
lower bound because we systematically overestimate connection at each
transformation. Using this transformation in Fig.~\ref{fig:block}(a),
we concluded $p_{c2} \ge 1/2$~\cite{bnd}, because the
limiting connection probability $r_{\infty}$ became one for $p \ge 1/2$.
Later in Ref.~\cite{hier}, it was pointed
out that the same method was applicable to find lower bounds for the usual
2D percolation thresholds as well, and also that such a lower bound
approached the true critical point $p_c$ as we used a larger block.
For the square lattice, a small block already predicts the correct critical
point of the bond percolation, $p^{\rm square}_c = 1/2$, and using a larger
block does not change this estimate, which can be an indication of the
exactness of $p^{\rm square}_c = 1/2$.
This observation motivates us to take a larger block in the EBT
[Fig.~\ref{fig:block}(b)], which requires us to check
$2^{23} \approx 8\times10^6$ configurations. The enumeration is
straightforward with a personal computer and the result is as follows:
\begin{eqnarray*}
r_{n+1} &=& 
9p^{12}r_n^7-91p^{11}r_n^7+409p^{10}r_n^7-1071p^9r_n^7+1795
 p^8r_n^7-1982p^7r_n^7+1414p^6r_n^7\\
&&-590p^5r_n^7+91p^4r_n^7+25p^3r_n^7-7p^2r_n^7-3pr_n^7+r_n^7-25p^{12}r_n^6+227p^{11}r_n^6\\
&&-904p^{10}r_n^6+2060p^9r_n^6-2928p^8r_n^6+2632p^7r_n^6-
1414p^6r_n^6+354p^5r_n^6+25p^4r_n^6\\
&&-27p^3r_n^6-2p^2r_n^6+2
pr_n^6+24p^{12}r_n^5-185p^{11}r_n^5+600p^{10}r_n^5-1038p^9
r_n^5+969p^8r_n^5\\
&&-354p^7r_n^5-165p^6r_n^5+210p^5r_n^5-57p^4r_n^5-9p^3r_n^5+5p^2r_n^5
-9p^{12}r_n^4+51p^{11}r_n^4\\
&&-102p^{10}r_n^4+60p^9r_n^4+60p^8r_n^4-70p^7r_n^4-46p^6r_n^4+92p
 ^5r_n^4-35p^4r_n^4-5p^3r_n^4\\
&&+4p^2r_n^4+p^{12}r_n^3-2p^{11}
r_n^3-24p^9r_n^3+115p^8r_n^3-198p^7r_n^3+150p^6r_n^3-32p^5
r_n^3\\
&&-18p^4r_n^3
+8p^3r_n^3-3p^{10}r_n^2+13p^9r_n^2-14p^8r_n^2-12p^7r_n^2+31p^6r_n^2-13p^5r_n^2\\
&&-6p^4r_n^2+4p^3r_n^2
+3 p^8r_n-16p^7r_n+31p^6r_n-26p^5r_n+8p^4r_n-p^6+5p^5\\
&&-8p^4+ 4p^3+p.
\end{eqnarray*}
The system-wide connection probability $r_{\infty}$ grows as we
increase $p$ [Fig.~\ref{fig:block}(c)].
By checking the value of $p$ where $r_{\infty}$ becomes $1$, we locate a
lower bound of $p_{c2}$. In fact, a careful look shows that $p=1/2$ is
still below $p_{c2}$ [Fig.~\ref{fig:block}(d)] and locates a sharper bound
$p_{c2} \gtrsim 0.523$.

\subsection{Transfer-matrix method}
\begin{figure}
\includegraphics[width=0.30\textwidth]{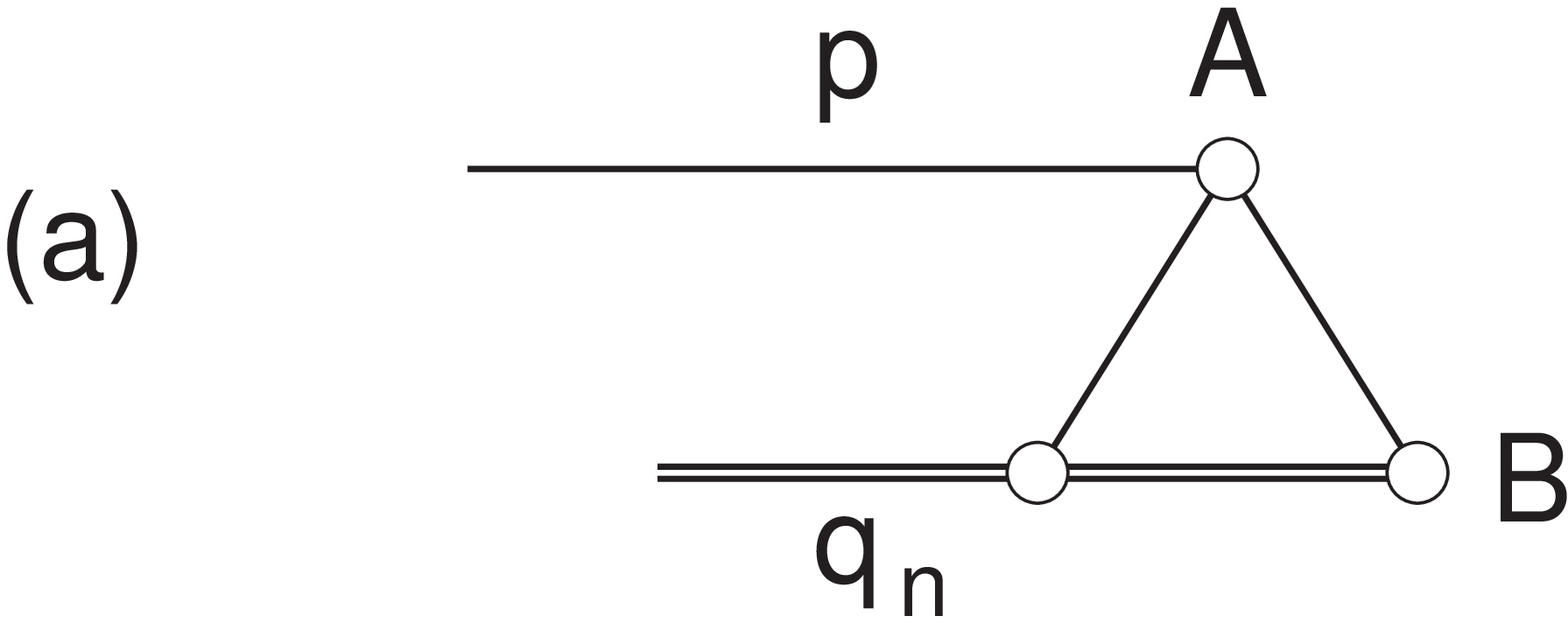}
\includegraphics[width=0.45\textwidth]{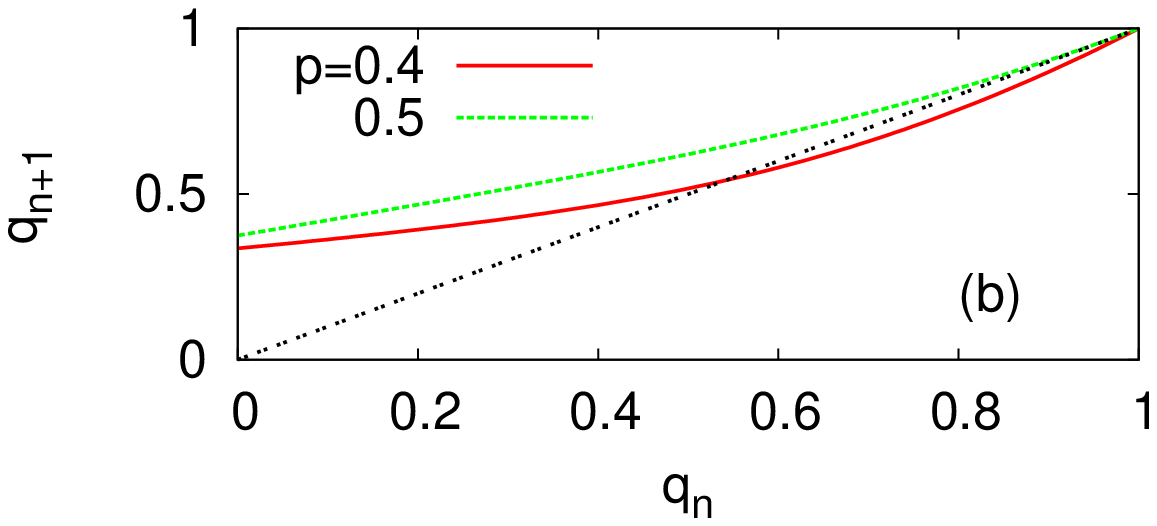}
\includegraphics[width=0.45\textwidth]{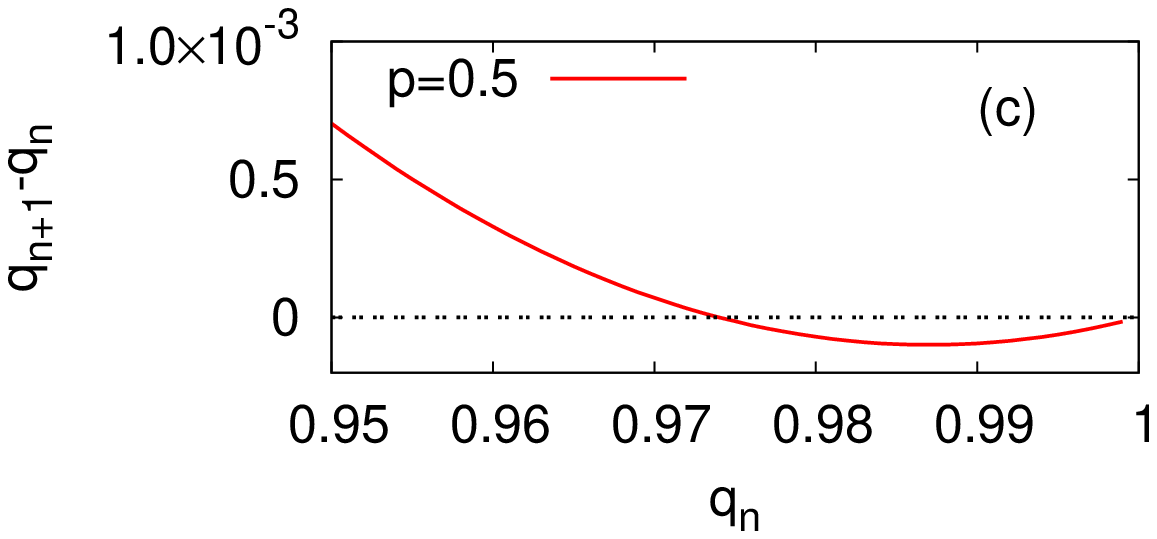}
\caption{(Color online)
(a) Unit cell for a transfer matrix calculation of a layer with
width 1. Each bond at the bottom line has an occupation probability $q_n$
(represented by a double line), while the other bonds have an occupation
probability $p$.
(b) Iteration maps for $p=0.4$ and $0.5$. (c) $p=0.5$ has an
intersection at $q_\infty < 1$ and therefore lies below $p_{c2}$. The
dotted lines indicate $q_{n+1} = q_n$.}
\label{fig:tm}
\end{figure}
In studying the Ising model on the EBT in Ref.~\cite{ising}, we
pointed out that the transfer-matrix method performed better than the
block-cell transformation. Hence, we employ the
transfer-matrix formalism developed for percolation in Ref.~\cite{derrida}.
First, we consider a unit cell of three spins as shown in
Fig.~\ref{fig:tm}(a). Note that a bond on the bottom line has a different
probability of $q_n$ from the others with $p$.
By attaching these cells from left to right, we
construct an indefinitely long strip, or a layer of width $1$,
which we can solve by using the transfer-matrix
method. When we consider connection to the leftmost side,
this cell has three possibilities: first, only the top point $A$ is
connected (case 1), second, only $B$ is connected (case 2), or finally, both
of them are connected (case 3). So we have nine possibilities of connection
in total as follows:
\begin{eqnarray*}
P_{11} \equiv P(1\rightarrow1) &=& p(p^2 q_n^2 - p q_n^2 - p q_n + 1),\\
P_{21} \equiv P(1\rightarrow2) &=& pq_n(1-p)(1+q_n-pq_n),\\
P_{31} \equiv P(1\rightarrow3) &=& p^2 q_n(1+q_n-pq_n),\\
P_{12} \equiv P(2\rightarrow1) &=& p^2 (1-q_n) (1+2q_n-2pq_n),\\
P_{22} \equiv P(2\rightarrow2) &=& q_n(1-p)(p^2+q_n+pq_n-2p^2q_n),\\
P_{32} \equiv P(2\rightarrow3) &=& p^2 q_n(p+2q_n-2pq_n),\\
P_{13} \equiv P(3\rightarrow1) &=& p(1-q_n)(1+q_n-pq_n),\\
P_{23} \equiv P(3\rightarrow2) &=& q_n(1-p)(p+q_n-pq_n),\\
P_{33} \equiv P(3\rightarrow3) &=& pq_n(p+q_n-pq_n).
\end{eqnarray*}
The global probability of connection to the leftmost side when $n(\gg 1)$
blocks are attached will behave as $\sim \lambda^n$, where $\lambda$ is the
largest eigenvalue of this $3\times3$ matrix $\{P_{ij}\}$. So we replace
this layer by a one-dimensional chain, and identify its occupation
probability $q_{n+1}$ with $\lambda$ to recover the original configuration
in Fig.~\ref{fig:tm}(a), but with $q_{n+1}$ instead of $q_n$.
This iteration therefore determines $q_{n+1}$ as
a function of $q_n$ and $p$, and we will find a limiting value $q_{\infty} =
\lim_{n\rightarrow \infty} q_n$ for a large system.
This renormalized connection probability is an
increasing function of $p$ [Fig.~\ref{fig:tm}(b)].
Again, we are interested in the value of $p$
making $q_{\infty}=1$, and such $p$ is found to be $\approx 0.504$.
In short, this confirms that $p=1/2$ is strictly
below $p_{c2}$ [Fig.~\ref{fig:tm}(c)].

\begin{figure}
\includegraphics[width=0.45\textwidth]{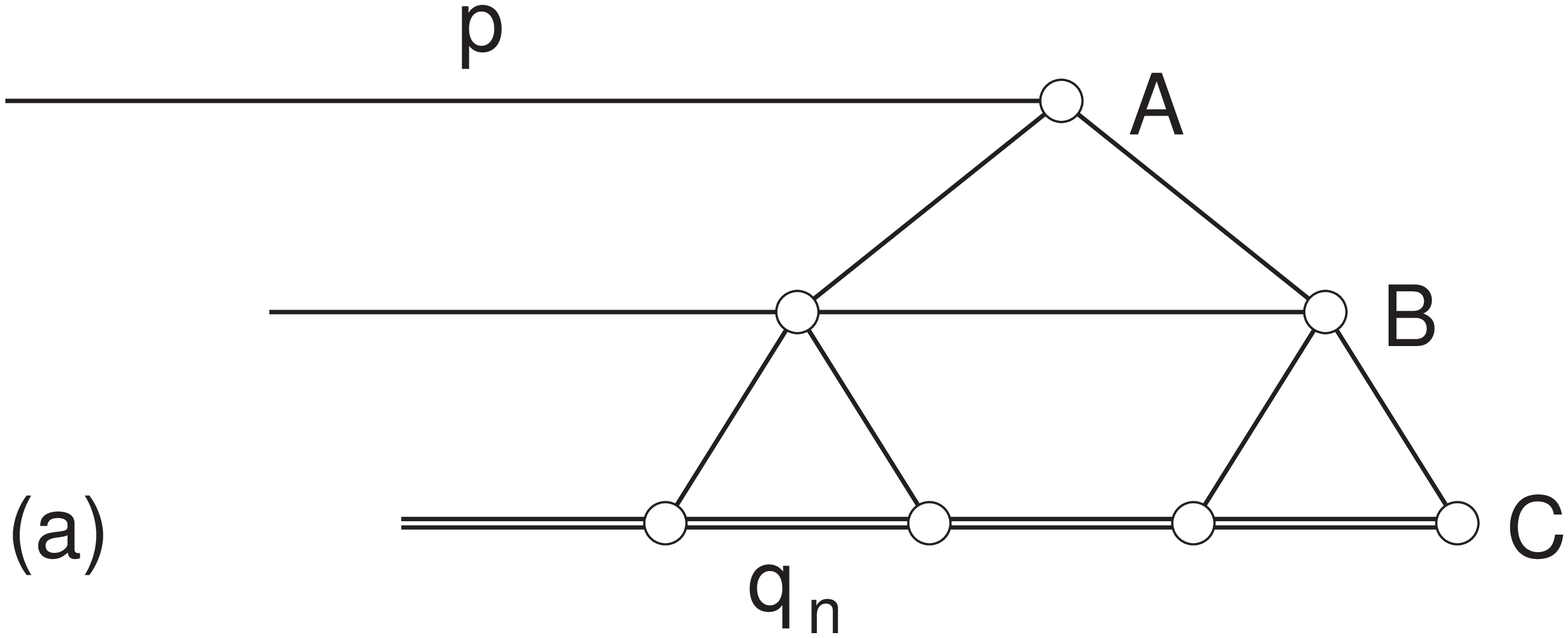}
\includegraphics[width=0.33\textwidth]{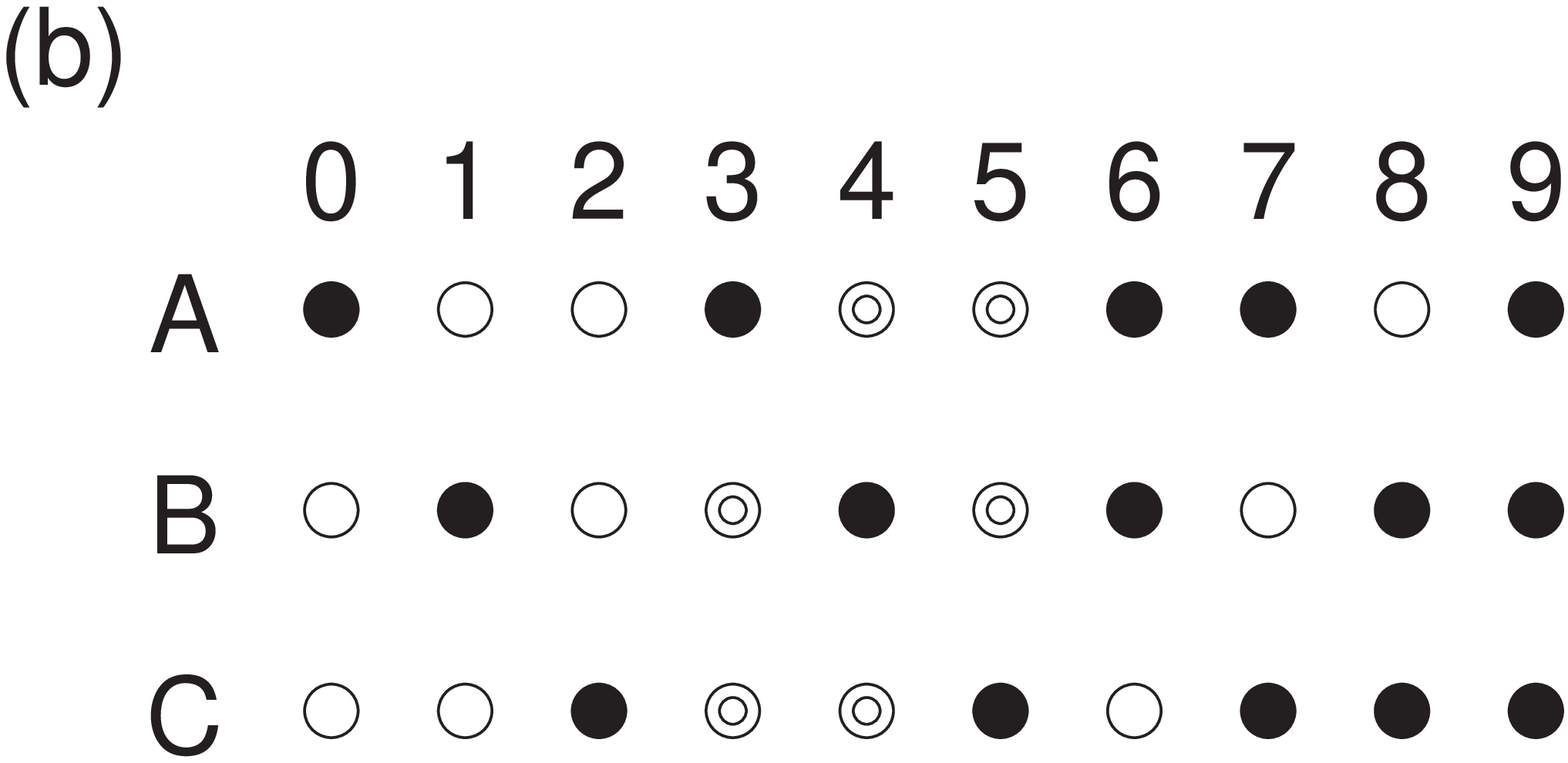}
\includegraphics[width=0.45\textwidth]{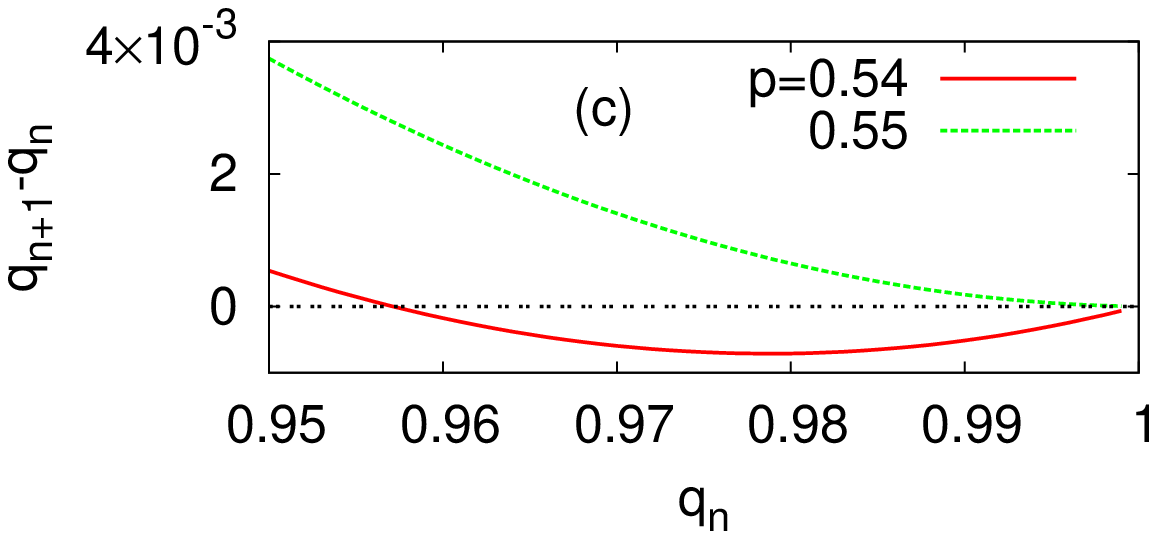}
\caption{(Color online)
(a) Unit cell for a transfer-matrix calculation of a layer with width $2$. (b)
Ten possible cases of connection. A black dot means that it is connected to
the leftmost side, and a white dot means it is not. A double circle means
that these
two are connected to each other, while none of them are connected to the
leftmost side. (c) The resulting iteration map suggests a new lower bound of
$p_{c2}$ as around $p=0.55$.}
\label{fig:tm2}
\end{figure}
We can take a larger layer, expecting a sharper bound
[Fig.~\ref{fig:tm2}(a)]. This consideration has ten possible cases as
listed in Fig.~\ref{fig:tm2}(b). Here, the black dots are connected to the
leftmost side while the white dots are not. It is important to consider
possibilities that two white dots may be connected to each other since a
percolating path may go backward for a while, so such dots are represented
by double circles.
In fact, the case indexed as $4$ is not
accessible from any other states so it can be discarded.
Each matrix element is expressed as a high-order polynomial, so
the average number of terms per polynomial amounts to $18.4$.
It is merely a
mechanical procedure to obtain the matrix elements so we show only the final
result with the largest eigenvalue, which is identified with $q_{n+1}$ as
above.
The result shown in Fig.~\ref{fig:tm2}(c) shows the highest
lower bound of $p_{c2}$, which is around $p=0.55$. It includes the numerical
lower bound suggested in Ref.~\cite{ziff} and the estimate based on the
duality relation~\cite{ebt}.

\section{Discussion}
\label{sec:con}

\begin{figure}
\includegraphics[width=0.45\textwidth]{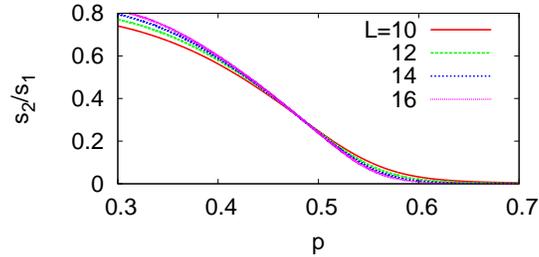}
\caption{(Color online)
Ratio of the second largest cluster size with respect to the largest
cluster size in the EBT with different numbers of layers $L$. Our calculation
suggests that the slope at $p=1/2$ will converge to a finite value
in the large-$N$ limit.}
\label{fig:ratio}
\end{figure}
At the time of writing Refs.~\cite{perc,comment}, we assumed that the
percolating properties in hyperbolic lattices could be inferred from known
2D results and also from the results of a tree.
For example, $s_2/s_1$ has a
diverging slope at the emergence of a giant cluster both for a 2D plane and for
a tree. That is why we expected the same behavior for hyperbolic structures
as well. However, as we see a clear difference from the 2D result in
Fig.~\ref{fig:tre}(b), such an assumption now looks quite dubious. Based on
the results so far obtained, it seems more plausible that this ratio also
has a constant slope at the large-$N$ limit (Fig.~\ref{fig:ratio}). This
implies that percolation in the EBT is neither similar to its 2D counterpart
nor to the percolation in a simple tree.
In particular, we see that the competition between the largest and the
second largest clusters appears milder than has been believed, so
that $s_2/s_1$ may vanish smoothly around $p_{c2}$.
In addition, it is inevitable to reconsider the phenomenological description
of the critical phenomena around $p_{c2}$ by using scaling
collapse~\cite{perc} since it is likely that
the critical points in the hyperbolic lattices
have been generally underestimated.
Our second example in Sec.~\ref{sec:solv}, however,
suggests that it can be difficult to extract the critical behavior if one
solely relies on numerical data. It will be interesting to challenge this
problem by making use of the recent analytic approaches to hierarchical
structures (see, e.g., Ref.~\cite{patch,boet}).

\acknowledgments
The author thanks Petter Minnhagen, Beom Jun Kim, and Robert M. Ziff
for their helpful comments.
This work was supported by the Supercomputing Center/Korea Institute of
Science and Technology Information with supercomputing resources including
technical support (Project No. KSC-2012-C1-05).


\begin{thebibliography}{18}
\expandafter\ifx\csname natexlab\endcsname\relax\def\natexlab#1{#1}\fi
\expandafter\ifx\csname bibnamefont\endcsname\relax
  \def\bibnamefont#1{#1}\fi
\expandafter\ifx\csname bibfnamefont\endcsname\relax
  \def\bibfnamefont#1{#1}\fi
\expandafter\ifx\csname citenamefont\endcsname\relax
  \def\citenamefont#1{#1}\fi
\expandafter\ifx\csname url\endcsname\relax
  \def\url#1{\texttt{#1}}\fi
\expandafter\ifx\csname urlprefix\endcsname\relax\def\urlprefix{URL }\fi
\providecommand{\bibinfo}[2]{#2}
\providecommand{\eprint}[2][]{\url{#2}}

\bibitem[{\citenamefont{Stauffer and Aharony}(1994)}]{stauffer}
\bibinfo{author}{\bibfnamefont{D.}~\bibnamefont{Stauffer}} \bibnamefont{and}
  \bibinfo{author}{\bibfnamefont{A.}~\bibnamefont{Aharony}},
  \emph{\bibinfo{title}{Introduction to Percolation Theory}},
  \bibinfo{edition}{2nd} ed.
  (\bibinfo{publisher}{CRC Press}, \bibinfo{address}{Boca Raton, FL},
  \bibinfo{year}{1994}).

\bibitem[{\citenamefont{Grimmett}(1999)}]{grimmett}
\bibinfo{author}{\bibfnamefont{G.}~\bibnamefont{Grimmett}},
  \emph{\bibinfo{title}{Percolation}}, \bibinfo{edition}{2nd} ed.
  (\bibinfo{publisher}{Springer},
  \bibinfo{address}{Berlin}, \bibinfo{year}{1999}).

\bibitem[{\citenamefont{Baek et~al.}(2009{\natexlab{a}})\citenamefont{Baek,
  Minnhagen, and Kim}}]{perc}
\bibinfo{author}{\bibfnamefont{S.~K.} \bibnamefont{Baek}},
  \bibinfo{author}{\bibfnamefont{P.}~\bibnamefont{Minnhagen}},
  \bibnamefont{and} \bibinfo{author}{\bibfnamefont{B.~J.} \bibnamefont{Kim}},
  \bibinfo{journal}{Phys. Rev. E} \textbf{\bibinfo{volume}{79}},
  \bibinfo{pages}{011124} (\bibinfo{year}{2009}{\natexlab{a}}).

\bibitem[{\citenamefont{Nogawa and Hasegawa}(2009)}]{ebt}
\bibinfo{author}{\bibfnamefont{T.}~\bibnamefont{Nogawa}} \bibnamefont{and}
  \bibinfo{author}{\bibfnamefont{T.}~\bibnamefont{Hasegawa}},
  \bibinfo{journal}{J. Phys. A} \textbf{\bibinfo{volume}{42}},
  \bibinfo{pages}{145001} (\bibinfo{year}{2009}).

\bibitem[{\citenamefont{Baek et~al.}(2009{\natexlab{b}})\citenamefont{Baek,
  Minnhagen, and Kim}}]{comment}
\bibinfo{author}{\bibfnamefont{S.~K.} \bibnamefont{Baek}},
  \bibinfo{author}{\bibfnamefont{P.}~\bibnamefont{Minnhagen}},
  \bibnamefont{and} \bibinfo{author}{\bibfnamefont{B.~J.} \bibnamefont{Kim}},
  \bibinfo{journal}{J. Phys. A} \textbf{\bibinfo{volume}{42}},
  \bibinfo{pages}{478001} (\bibinfo{year}{2009}{\natexlab{b}}).

\bibitem[{\citenamefont{Minnhagen and Baek}(2010)}]{an}
\bibinfo{author}{\bibfnamefont{P.}~\bibnamefont{Minnhagen}} \bibnamefont{and}
  \bibinfo{author}{\bibfnamefont{S.~K.} \bibnamefont{Baek}},
  \bibinfo{journal}{Phys. Rev. E} \textbf{\bibinfo{volume}{82}},
  \bibinfo{pages}{011113} (\bibinfo{year}{2010}).

\bibitem[{\citenamefont{Baek and Minnhagen}(2011{\natexlab{a}})}]{bnd}
\bibinfo{author}{\bibfnamefont{S.~K.} \bibnamefont{Baek}} \bibnamefont{and}
  \bibinfo{author}{\bibfnamefont{P.}~\bibnamefont{Minnhagen}},
  \bibinfo{journal}{Physica A} \textbf{\bibinfo{volume}{390}},
  \bibinfo{pages}{1447} (\bibinfo{year}{2011}{\natexlab{a}}).

\bibitem[{\citenamefont{Gu and Ziff}()}]{ziff}
\bibinfo{author}{\bibfnamefont{H.}~\bibnamefont{Gu}} \bibnamefont{and}
  \bibinfo{author}{\bibfnamefont{R.~M.} \bibnamefont{Ziff}},
  \bibinfo{note}{arXiv:1111.5626}.

\bibitem[{\citenamefont{Cardy}(1992)}]{cardy}
\bibinfo{author}{\bibfnamefont{J.~L.} \bibnamefont{Cardy}},
  \bibinfo{journal}{J. Phys. A} \textbf{\bibinfo{volume}{25}},
  \bibinfo{pages}{L201} (\bibinfo{year}{1992}).

\bibitem[{\citenamefont{Kleban and Vassileva}(1994)}]{kleban}
\bibinfo{author}{\bibfnamefont{P.}~\bibnamefont{Kleban}} \bibnamefont{and}
  \bibinfo{author}{\bibfnamefont{I.}~\bibnamefont{Vassileva}},
  \bibinfo{journal}{Phys. Rev. Lett.} \textbf{\bibinfo{volume}{72}},
  \bibinfo{pages}{3929} (\bibinfo{year}{1994}).

\bibitem[{\citenamefont{Rietman et~al.}(1992)\citenamefont{Rietman, Nienhuis,
  and Otimaa}}]{rietman}
\bibinfo{author}{\bibfnamefont{R.}~\bibnamefont{Rietman}},
  \bibinfo{author}{\bibfnamefont{B.}~\bibnamefont{Nienhuis}}, \bibnamefont{and}
  \bibinfo{author}{\bibfnamefont{J.}~\bibnamefont{Otimaa}},
  \bibinfo{journal}{J. Phys. A} \textbf{\bibinfo{volume}{25}},
  \bibinfo{pages}{6577} (\bibinfo{year}{1992}).

\bibitem[{\citenamefont{Williams}(1991)}]{will}
\bibinfo{author}{\bibfnamefont{D.}~\bibnamefont{Williams}},
  \emph{\bibinfo{title}{Probability with Martingales}}
  (\bibinfo{publisher}{Cambridge University Press},
  \bibinfo{address}{Cambridge}, \bibinfo{year}{1991}).

\bibitem[{\citenamefont{Ziff}(2006)}]{cell}
\bibinfo{author}{\bibfnamefont{R.~M.} \bibnamefont{Ziff}},
  \bibinfo{journal}{Phys. Rev. E} \textbf{\bibinfo{volume}{73}},
  \bibinfo{pages}{016134} (\bibinfo{year}{2006}).

\bibitem[{\citenamefont{Baek and Minnhagen}(2011{\natexlab{b}})}]{hier}
\bibinfo{author}{\bibfnamefont{S.~K.} \bibnamefont{Baek}} \bibnamefont{and}
  \bibinfo{author}{\bibfnamefont{P.}~\bibnamefont{Minnhagen}},
  \bibinfo{journal}{Phys. Scr.} \textbf{\bibinfo{volume}{83}},
  \bibinfo{pages}{055601} (\bibinfo{year}{2011}{\natexlab{b}}).

\bibitem[{\citenamefont{Baek et~al.}(2011)\citenamefont{Baek, M{\"a}kel{\"a},
  Minnhagen, and Kim}}]{ising}
\bibinfo{author}{\bibfnamefont{S.~K.} \bibnamefont{Baek}},
  \bibinfo{author}{\bibfnamefont{H.}~\bibnamefont{M{\"a}kel{\"a}}},
  \bibinfo{author}{\bibfnamefont{P.}~\bibnamefont{Minnhagen}},
  \bibnamefont{and} \bibinfo{author}{\bibfnamefont{B.~J.} \bibnamefont{Kim}},
  \bibinfo{journal}{Phys. Rev. E} \textbf{\bibinfo{volume}{84}},
  \bibinfo{pages}{032103} (\bibinfo{year}{2011}).

\bibitem[{\citenamefont{Derrida and Vannimenus}(1980)}]{derrida}
\bibinfo{author}{\bibfnamefont{B.}~\bibnamefont{Derrida}} \bibnamefont{and}
  \bibinfo{author}{\bibfnamefont{J.}~\bibnamefont{Vannimenus}},
  \bibinfo{journal}{J. Phys. (France) Lett.} \textbf{\bibinfo{volume}{41}},
  \bibinfo{pages}{L473} (\bibinfo{year}{1980}).

\bibitem[{\citenamefont{Boettcher et~al.}(2009)\citenamefont{Boettcher, Cook,
  and Ziff}}]{patch}
\bibinfo{author}{\bibfnamefont{S.}~\bibnamefont{Boettcher}},
  \bibinfo{author}{\bibfnamefont{J.~L.} \bibnamefont{Cook}}, \bibnamefont{and}
  \bibinfo{author}{\bibfnamefont{R.~M.} \bibnamefont{Ziff}},
  \bibinfo{journal}{Phys. Rev. E} \textbf{\bibinfo{volume}{80}},
  \bibinfo{pages}{041115} (\bibinfo{year}{2009}).

\bibitem[{\citenamefont{Boettcher and Brunson}(2011)}]{boet}
\bibinfo{author}{\bibfnamefont{S.}~\bibnamefont{Boettcher}} \bibnamefont{and}
  \bibinfo{author}{\bibfnamefont{C.~T.} \bibnamefont{Brunson}},
  \bibinfo{journal}{Phys. Rev. E} \textbf{\bibinfo{volume}{83}},
  \bibinfo{pages}{021103} (\bibinfo{year}{2011}).

\end{thebibliography}

\end{document}